\documentclass[%
 amsmath,amssymb,
 preprint,%
 prl,
 superscriptaddress,
 floatfix
]{revtex4-2}

\usepackage{graphicx}
\usepackage{dcolumn}
\usepackage{bm}
\usepackage{ulem}
\usepackage[colorlinks,linkcolor=blue]{hyperref}
\usepackage{mathrsfs}
\usepackage{verbatim}
\usepackage{siunitx}

\begin{document}

\preprint{AIP/123-QED}
\title[Jetting from resonant inertial cavitation]{Amplification of supersonic micro-jets by resonant inertial cavitation-bubble pair} 
\author{Yuzhe Fan}
\thanks{Yuzhe Fan and Alexander Bußmann are equal contributors to this work and designated as co-first authors.}
\affiliation{Faculty of Natural Sciences, Institute for Physics, Department Soft Matter, Otto-von-Guericke University Magdeburg,39106 Magdeburg, Germany}
\affiliation{Research Campus STIMULATE, University of Magdeburg, Otto-Hahn-Straße 2, 39106 Magdeburg, Germany}
\author{Alexander Bußmann}
\thanks{Yuzhe Fan and Alexander Bußmann are equal contributors to this work and designated as co-first authors.}
\affiliation{Chair of Aerodynamics and Fluid Mechanics, TUM School of Engineering and Design, Technical University of Munich, 85748 Garching bei München, Germany}
\affiliation{Munich Institute of Integrated Materials, Energy and Process Engineering (MEP), Technical University of Munich, 85748 Garching bei München, Germany}
\author{Fabian Reuter}
\affiliation{Faculty of Natural Sciences, Institute for Physics, Department Soft Matter, Otto-von-Guericke University Magdeburg,39106 Magdeburg, Germany}
\author{Hengzhu Bao}
\affiliation{Suzhou University of Science and Technology, School of Physical Science and Technology, Suzhou 215009, China}
\author{Stefan Adami}
\affiliation{Munich Institute of Integrated Materials, Energy and Process Engineering (MEP), Technical University of Munich, 85748 Garching bei München, Germany}
\author{José M. Gordillo}
\affiliation{Área de Mecánica de Fluidos, Departamento de Ingenería Aeroespecial y Mecánica de Fluidos, Universidad de Sevilla, 41092 Sevilla, Spain}
\author{Nikolaus Adams}
\affiliation{Chair of Aerodynamics and Fluid Mechanics, TUM School of Engineering and Design, Technical University of Munich, 85748 Garching bei München, Germany}
\affiliation{Munich Institute of Integrated Materials, Energy and Process Engineering (MEP), Technical University of Munich, 85748 Garching bei München, Germany}
\author{Claus-Dieter Ohl}
\affiliation{Faculty of Natural Sciences, Institute for Physics, Department Soft Matter, Otto-von-Guericke University Magdeburg,39106 Magdeburg, Germany}
\affiliation{Research Campus STIMULATE, University of Magdeburg, Otto-Hahn-Straße 2, 39106 Magdeburg, Germany}
\date{\today}
\begin{abstract}
We reveal for the first time by experiments that within a narrow parameter regime, two cavitation bubbles with identical energy generated in anti-phase develop a supersonic jet. High-resolution numerical simulation shows a mechanism for jet amplification based on toroidal shock wave and bubble necking interaction. The micro-jet reaches velocities in excess of $\SI{1000}{\meter\per\second}$. We demonstrate that potential flow approximation established for Worthington jets accurately predicts the evolution of the bubble gas-liquid interfaces.

\end{abstract}

\maketitle

\paragraph{Introduction:} 
Liquid jets are encountered in various physical scenarios of collapsing cavities in liquids. Often jets are initiated by the collapse of conical or axisymmetric parabolic cavities \cite{gordillo2023jets, Eggers2007}, e.g., by bubbles bursting near a free surface \cite{GananCalvo2017, Lai2018, CommentBursting,Ji2021,gordillo2023jets}, collapsing depressions of standing waves \cite{Zeff2000}, the impact of water droplets on hydrophobic surfaces \cite{bartolo2006singular}, or the impact of solids on a free surface \cite{Bergmann2006, gekle2009high, gekle2010supersonic}. In those scenarios, the cavity collapse leads to impressive amplifications of the jet velocity compared to their initiation mechanism, where the jet velocity varies from low subsonic \cite{bartolo2006singular, Bergmann2006, gekle2009high, Zeff2000} to supersonic states \cite{gekle2010supersonic}. The enhancement of such energy-focusing mechanisms through the utilization of single/multiple cavities is of great importance, e.g., for alternative medical therapies of needle-based injections \cite{Schoppink2022} or soft material perforation \cite{robles2020soft}, where stable jets with high velocities and small volume are desired. Recently, high-speed liquid jets have been found during the evolution of single cavitation bubbles close to boundaries \cite{lechner2019fast,reuter2021supersonic, Bussmann2023, sieber2023cavitation}, where radial inward rushing liquid toward the symmetry axis induces a singularity and corresponding supersonic needle jets. However, the existence of supersonic jets during the interaction of multiple cavitation bubbles has not been reported so far.

\begin{figure*}
  \centering
  \includegraphics[width=0.98\textwidth]{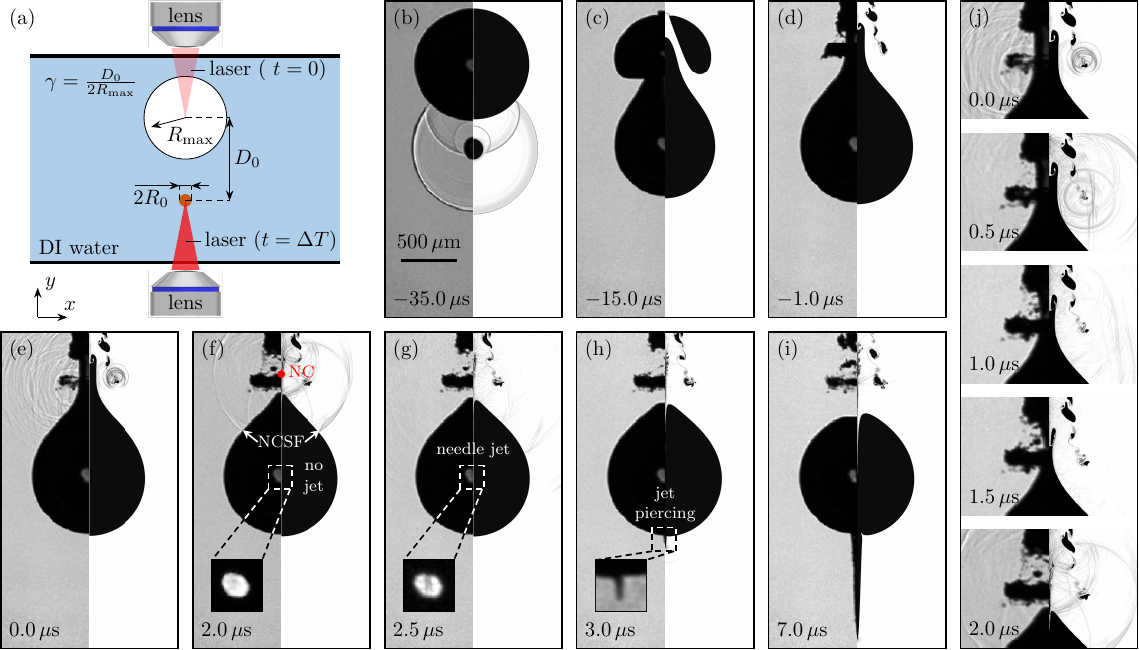}
  \caption{\textbf{High-speed images and comparison between simulation and experiment in the ultra-fast needle jet regime for a non-dimensional standoff distance of $\gamma\approx0.75$.} (a) Schematic representation of the camera view of the experimental set-up and the initial bubble sizes in the simulation. The upper bubble is generated first, and after $t=\Delta T\approx\SI{46}{\micro\second}$, the second lower bubble is initiated. All lengths are not drawn to scale. (b-j) High-speed side-view dynamics of the experiment (left half of each image) and visualization of numerical Schlieren in the simulation (right half of each image). Note that experimental shadowgraphs show a bubble projection while the simulation shows a plane cutting the bubble, revealing more details of the bubble shape. Time instances are given in the lower corner of each image, where $t=\SI{0}{\micro\second}$ corresponds to the time of the torus collapse of the first bubble. In (j) a detailed view on the neck collapse between (e) and (f) is shown. In (f) the red dot marks the position of the neck collapse on the symmetry axis (NC), and the white arrows indicate the position of the neck-collapse shock front (NCSF).}
  \label{fig:illustration}
\end{figure*}

To reveal the interaction of bubbles experimentally, we reduce the complexity to the most simple problem and study the dynamics of a single bubble pair. These have been investigated previously on different small and large scales \cite{sankin2010pulsating,yuan2011dynamics,robles2020soft, zhang2016experimental}. \citet{han2015dynamics} systematically compared experiments with simulations by varying the distance and the time interval between bubble generation. They and others have shown that the generation of similarly sized bubble pairs in anti-phase, i.e., the second bubble is generated when the first bubble is at maximum size, results in fast jets  \cite{cui2016experimental,luo2019jet,chew2011interaction,liang2021interaction,han2015dynamics,tomita2017pulsed,robles2020soft,yuan2011dynamics, Mishra2022}, where jet velocities inside the cavitation bubble of $O(100)\,\si{\meter\per\second}$ are reached \cite{han2015dynamics, tomita2017pulsed, Mishra2022}. Focusing on the identical anti-phase bubble pair situation, in this Letter, we demonstrate experimentally and numerically that at a proper distance, the bubble pair can accelerate a needle-type jet an order of magnitude faster than formerly reported. It resembles the needle jet found in cavitation erosion studies and at elastic boundaries, yet it does not atomize but stably penetrate the gas-liquid interface. 

\paragraph{Experimental and numerical setup:} 
In this study, we focus on the anti-phase dynamics of bubble pairs with identical maximum radius, $R_{1,\mathrm{max}}=R_{2,\mathrm{max}}=R_{\mathrm{max}}$. Here, $R_{i,\mathrm{max}}$ are the maximum radii each bubble would reach independently, i.e., without the neighboring bubble. We control anti-phase dynamics by initiating the second bubble at $\Delta T = T_{1,\mathrm{max}} = T_{2,\mathrm{max}} = T_{\mathrm{max}}$, where $T_{i,\mathrm{max}}$ is the time when the bubble reaches its maximum size. The remaining variable parameter is the bubble seeding distance $D_0$, which reads in non-dimensional form $\gamma = D_0/(2R_{\mathrm{max}})$.

In the experiments, two bubbles are generated in deionized (DI) water through optical breakdown from two pulsed lasers (Litron Nano T-250-10; wavelength $\SI{1064}{\nano\meter}$; pulse duration $\mathrm{FWHM}=\SI{7}{\nano\second}$, and Litron Nano SG-100-2; $\SI{1064}{\nano\meter}$; $\mathrm{FWHM}=\SI{6}{\nano\second}$). The water is contained in a cuboid optical glass cuvette with squared cross section of side length of $\SI{2}{\centi\meter}$ and a height of $\SI{5}{\centi\meter}$ (Fig.~\ref{fig:illustration}a). Two long working distance microscope objectives facing each other focus the laser pulses. The objective lenses are mounted on three-axis micrometer stages to align the foci in the imaging plane and control the distance between the two bubbles. The dynamics of the bubbles are recorded with a high-speed camera (Shimadzu XPV-X2) at 2 million frames per second equipped with a macro lens (MP-E $\SI{65}{\milli\meter}$ f/2.8 1-5x Macro) with its magnification set to a fixed pixel resolution of $\SI{7}{\micro\meter}$ per pixel. The effective exposure time is $\SI{220}{\femto\second}$ using an expanded pulse train from a femtosecond laser beam that illuminates the scene (Ekspla Femtolux, wavelength $\SI{515}{\nano\meter}$).

We compare the experimental bubble dynamics with high-resolution direct numerical simulations. We employ the open-source code framework ALPACA \cite{ALPACA, Hoppe2022_ALPACA, Hoppe2022_levelset}, which solves the compressible three-dimensional Navier-Stokes equations, including viscous and capillary effects, by a Godunov-type flux-based high-order finite-volume formulation. We neglect species diffusion, gravity, and heat exchange. A conservative sharp-interface level-set method captures the interface between the cavitation bubble and the surrounding liquid. Computational costs are reduced by employing a multi-resolution scheme with timestep adaptivity and a three-dimensional axisymmetric domain. We model the bubble content by a single non-condensable ideal gas, and the surrounding water follows the Tait equation of state \cite{Bussmann2023}. Bubble dynamics are represented by initial high-energetic spherical bubbles ($R_0\approx\SI{20}{\micro\meter}$, $p_0=\SI{11000}{\bar}$) that expand into water at rest (Fig.~\ref{fig:illustration}a). We obtain initial conditions for every initiation by matching the dynamics of a single cavitation bubble in the experiments (see supplementary material (SM)). Phase-change effects are taken into account by reducing the equilibrium radius of the first initiated bubble by 65 \% at its maximum \cite{koch2016numerical,liang2022comprehensive}. The mesh size in the simulation is sufficient to resolve the initial bubble with approximately $N_\mathrm{min}\approx50$ cells, and shock wave reflections at the domain boundaries are suppressed by choosing the domain sufficiently large ($L=90R_\mathrm{max}$).

\paragraph{Results and discussion:} Figure~\ref{fig:illustration}b-j gives an overview of the dynamics of an anti-phase bubble pair with $\gamma \approx 0.75$, where the upper/lower bubble is initiated first/second. Composite images showing the experiment on the left and the simulation on the right half of each tile. 
The overall time scales of the numerical prediction were slightly different, which we associate with the simplified representation of the bubble material by a single ideal gas. Additionally, the experimental bubble appears to be slightly elongated shortly after breakdown, whereas for the simulation we initiate a spherical bubble. To compensate for the slight time-scale difference between experiment and simulation, frames of composite images are chosen in time such that both exhibit the same interface dynamics and thus correspond to the same stage of dynamics. We observe excellent agreement between simulation and experiment for the bubble dynamics and shock wave emission. Figure~\ref{fig:illustration}b shows images just after the initiation of the second bubble, where the plasma shock wave and its reflections are visible. While the first bubble is already shrinking, the second grows into the first, forming a long neck. This neck deforms the first bubble into a torus passing through the torus center, see Fig.~\ref{fig:illustration}c-e. Eventually, the first bubble collapses (Fig.~\ref{fig:illustration}e) and the neck of the second bubble breaks up (Fig.~\ref{fig:illustration}f). Immediately after neck collapse, a fast jet is ejected, as can be seen in the bubble interior, inset of Fig.~\ref{fig:illustration}g. Thus, it has already passed more than half of the bubble within $\SI{500}{\nano\second}$. It pierces the lower bubble wall (Fig.~\ref{fig:illustration}h), and further grows in length and radius (Fig.~\ref{fig:illustration}i). 

Neck formation and collapse are key events in the dynamics leading to the ultra-fast needle jet. Neck formation can be seen in Fig.~\ref{fig:illustration}c-e. Collapse, jet formation, and piercing of the jet through the distal part of the second bubble are shown in Fig.~\ref{fig:illustration}g-j. The average jet velocity can be measured from the instant of formation to the instant when it is seen to have passed the bubble center. It exhibits an average velocity of $\SI{1400}{\meter\per\second}$, where from the simulations, peak velocities up to $\SI{3000}{\meter\per\second}$ can be measured immediately after jet ejection, which reduce to $\SI{1200}{\meter\per\second}$ before the jet impacts the distal side of the second bubble. After the jet has penetrated the second bubble, it remains at a reduced speed of $\SI{250}{\meter\per\second}$. 

In Fig.~\ref{fig:illustration}j, a detailed description with finer temporal resolution of the neck breakup is shown. Shortly before the collapse of the first bubble, the neck resembles a cylindrical shape reaching its maximum elongation. Once the torus collapses ($t=\SI{0}{\micro\second}$), it emits a shock wave towards the elongated neck. After the shock wave emission, the neck starts to contract with a parabolic shape towards the axis of symmetry without reducing its axial neck length ($\SI{0}{\micro\second} \leq t \leq \SI{1.5}{\micro\second}$). It converges radially on the symmetry axis, which results in the emission of a strong shock wave, visible in the experiment and simulation (see red dot and white arrows in Fig.~\ref{fig:illustration}f). 

\begin{figure}
  \centering
  \includegraphics[width=0.48\textwidth]{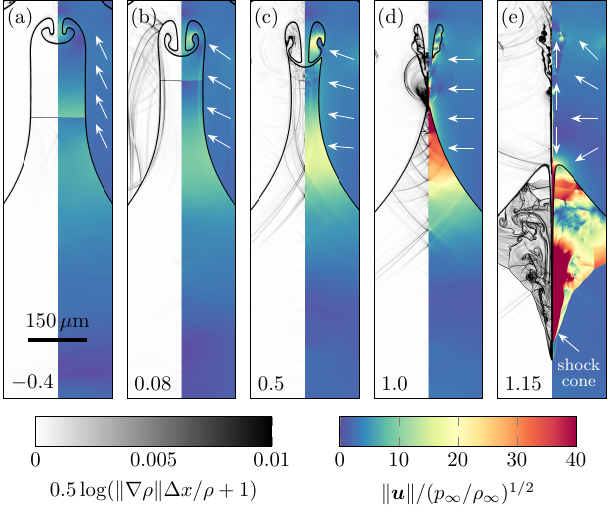}
  \caption{\textbf{Detailed numerical representation of the ultra-fast needle jet generation dynamics shortly before and after the torus collapse.} Non-dimensional time instances are given in the lower left corner of each image, where $0$ corresponds to the time of torus collapse and $1$ to the time of neck breakup. The left half of each image shows non-dimensional numerical Schlieren and the right half shows the normalized velocity magnitude, where $(p_\infty/\rho_\infty)^{1/2}\approx\SI{10}{\meter\per\second}$. The velocity field is overlaid by non-scaled white arrows indicating the flow direction. Colorbars for all images are represented at the lower part of the image.}
  \label{fig:neck_evolution_sim}
\end{figure}

The following illustrations give detailed insight into the neck breakup and jet formation mechanisms. We use a non-dimensional time, $t_\ast=(t-t_\mathrm{tc})/(t_\mathrm{nc}-t_\mathrm{tc})$, where $t_\mathrm{tc}$ is the time of the torus collapse of the first bubble and $t_\mathrm{nc}$ the breakup of the elongated neck. Hence, negative/positive values represent times before and after the torus collapse, respectively. The instant of neck breakup is the time when the radial flow following the collapsing neck converges on the axis of symmetry, which is detected by shock wave emission (Fig.~\ref{fig:neck_evolution_sim}d). Hence, the neck breakup is given by $t_\ast=1$. 

In Fig.~\ref{fig:neck_evolution_sim}, a detailed description with finer temporal resolution of the neck breakup is shown. We observe that the liquid flow velocity has a component tangential to the neck interface before the torus bubble collapses (Fig.~\ref{fig:neck_evolution_sim}a). Hence, only a small radial contraction is observed. Once the torus bubble has collapsed, the emitted toroidal shock wave impacts on the neck. This wave is partly reflected from and transmitted into the neck (Fig.~\ref{fig:neck_evolution_sim}b). The result is a contraction of the neck with a parabolic shape enhancing the radial inward motion of the surrounding liquid (Fig.~\ref{fig:neck_evolution_sim}b-d). Once the neck has contracted on the symmetry axis strong water-hammer shock waves are emitted (Fig.~\ref{fig:neck_evolution_sim}d). Finally, this high pressure region drives the flow upward/downward in the axial direction, forming the ultra-fast needle jet (Fig.~\ref{fig:neck_evolution_sim}e). The supersonic velocity of the collapsing neck causes a conical shock wave to form inside the bubble shortly after the neck breakup (Fig.~\ref{fig:neck_evolution_sim}e). The shock cone half-angle is measured as 16 degrees, implying a jet velocity of about 3.6 times the speed of sound of the bubble content.

\begin{figure}
  \centering
  \includegraphics[width=0.48\textwidth]{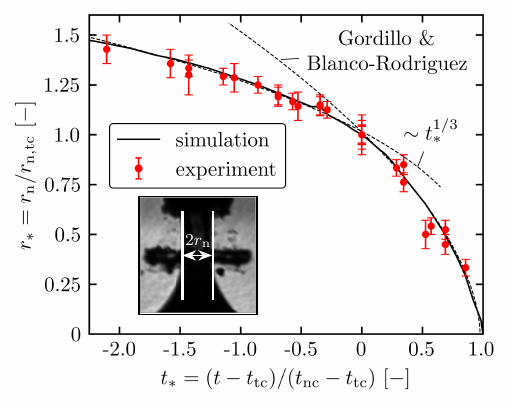}
  \caption{\textbf{Temporal evolution of the neck radius.} Non-dimensional neck radius, $r^\ast = r_\mathrm{n}/r_\mathrm{n, tc}$, as function of the non-dimensional time, $t^\ast = (t_\mathrm{nc} - t)/(t_\mathrm{nc} - t_\mathrm{tc})$. Here, $r_\mathrm{n, tc}$ is the neck radius at the time of torus collapse, $t_\mathrm{tc}$, and $t_\mathrm{nc}$ the time of neck collapse. The measurement of the neck radius, $r_\mathrm{n}$, is indicated in the image in the lower left corner. Experimental data are from different experiments within the $\gamma\in[0.72, 0.79]$ range. The line for the simulation represents a standoff distance of $\gamma\approx0.75$. Additionally, an analytical solution of \citet{gordillo2023jets} for the collapse of a parabolic cavity is shown (see SM for further details).}
  \label{fig:neck_radius_over_time}
\end{figure}

To further explain the mechanisms inducing the ultra-fast needle jet, Fig.~\ref{fig:neck_radius_over_time} shows the temporal evolution of the non-dimensional neck radius, $r_\mathrm{n}/r_\mathrm{n,tc}$, where $r_\mathrm{n,tc}$ is the neck radius at torus collapse. The experimental data are given for different non-dimensional standoff distances within the $\gamma\in[0.72, 0.79]$ range. 
We identify two regimes: a weak contraction regime, $t_\ast<0$, where the neck reduces proportional by to $t_\ast^{1/3}$, and a shock-accelerated regime, $t_\ast>0$. Before the torus collapse, the flow field is partly tangential to the neck interface, reducing the radial contraction rate of the neck (Fig.~\ref{fig:neck_evolution_sim}a). After the impact of the torus shock wave and the impulsive acceleration of the neck, the neck radius reduces with an increased averaged slope, which agrees well with the analytical model in \cite{gordillo2023jetsSupplementary} and references therein for the collapse of a parabolic cavity.

\begin{figure}
  \centering
  \includegraphics[width=0.48\textwidth]{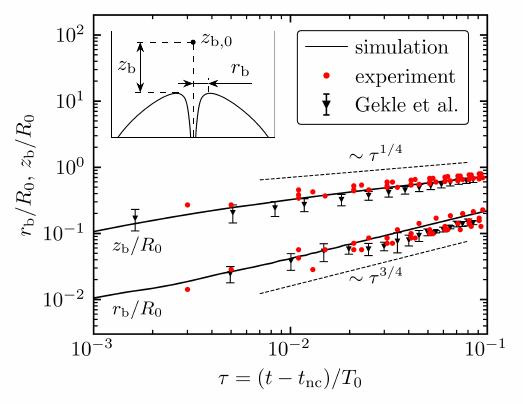}
  \caption{\textbf{Temporal evolution of the base of the ultra-fast needle jet.} Non-dimensional jet base coordinates, $r_\mathrm{b}/R_0$, and, $z_\mathrm{b}/R_0$, as function of the non-dimensional time, $\tau = (t-t_\mathrm{nc})/T_0$. Here, $t_\mathrm{nc}$ is the time of neck collapse, $R_0=R_\mathrm{max}\approx\SI{500}{\micro\meter}$ the reference length, and $T_0=R_\mathrm{max}(\rho_\infty/p_\infty)^{1/2}\approx\SI{50}{\micro\second}$ the reference time. A schematic sketch of the jet base geometry is given in the upper left corner, where $z_{\mathrm{b},0}$ is the axial position of the jet pinch-off. Experimental data are taken from different experiments within the $\gamma\in[0.72, 0.79]$ range, and we estimate the uncertainty for $r_\mathrm{b}/R_0$ as well as $z_\mathrm{b}/R_0$ as 0.02. 
  The line for the simulation represents a standoff distance of $\gamma\approx0.75$. Experimental data of \citet{gekle2009high} are also added.}
  \label{fig:jet_base_over_time}
\end{figure}

As initial conditions, the simulated values of the neck radius, $r_\mathrm{n}$, the neck velocity, $\dot{r}_\mathrm{n}$, and the radius of curvature of the neck, $r_\mathrm{c}$, are taken at $t_\ast\approx0.5$, where the inflow on the neck is mainly radial (Fig.~\ref{fig:neck_evolution_sim}c). The results in Fig.~\ref{fig:neck_radius_over_time} and in the SM show that the radial flow rate per unit length, induced at the instant when the torus collapses, decreases very slowly (logarithmically), fixing the value of the flow rate per unit length, $q_\infty$, driving the ejection of the jet \cite{gordillo2023jets}. Indeed, \citet{gordillo2023jets} show that the time evolutions of the jet width, $r_\mathrm{b}$, and the jet vertical position, $z_\mathrm{b}$, after emission from an implosion of parabolic cavities are respectively given by $r_\mathrm{b}\propto \left(q^3_\infty(t-t_\mathrm{nc})^3/r^2_\mathrm{c}\right)^{1/4}$, and $z_\mathrm{b}\propto \left(q_\infty(t-t_\mathrm{nc})r^2_\mathrm{c}\right)^{1/4}$. These predictions are confirmed by the experimental and numerical results depicted in Fig.~\ref{fig:jet_base_over_time}, which exhibit the same dependence with time after a short transient.
Interestingly, the results in Fig.~\ref{fig:jet_base_over_time} also reveal that Worthington jets produced by the incompressible collapse of parabolic cavities induced after solid impact on a free surface \cite{gekle2009high, gekle2010jfm} show the same time dependence as the compressible jets produced after the cavitation of bubbles. This finding can be explained by the fact that the velocity field created by an acoustic monopole at a distance $r\ll a_\infty (t-t_\mathrm{nc})$, where $a_\infty$ represents the speed of sound in the liquid, closely resembles the one of an incompressible sink.

We observe the formation of the ultra-fast needle jet only in a narrow range of standoff distances, $\gamma\in[0.7,0.8]$ (see SM). For larger standoff distances, $\gamma>0.8$, the elongated neck of the second bubble does not pierce the first bubble, which prevents the impact of the torus shock wave on the neck and, consequently, a radial neck contraction. For smaller standoff distances, $\gamma<0.7$, the neck grows in length and size, favoring unstable neck breakup with multiple breakup points on the neck surface at some radial distance to the axis of symmetry. Hence, we identify that the collapse of a parabolic cavity dominates the formation of the needle jet for equal-sized anti-phase bubble pairs.

\paragraph{Summary:} Based on the observations of high-speed imaging experiments and high-resolution numerical simulations, we have revealed that the dynamics of a laser-generated anti-phase bubble pair can form ultra-fast needle jets with peak velocities up to $\SI{3000}{\meter\per\second}$ reached immediately after jet ejection and reduced to $\SI{1200}{\meter\per\second}$ before the jet impacts the distal side of the second bubble. The jet remains stable after its appearance on the distal part of the second bubble, and the jet base evolution agrees well with those observed for Worthington jets. Such stable supersonic jets only emerge in a narrow range of standoff distances between the two bubbles. Only then, a singularity forms on the axis of symmetry, created by the collapse of the elongated neck of the second bubble with a parabolic shape, which is necessary for the ultra-fast needle jet formation. Shock waves emitted from the toroidal collapse of the first bubble converge towards the neck, trigger parabolic neck collapse, and further amplify jet velocity. Due to their high velocity but small size, such supersonic jets may improve technical applications, such as needle-free injections. From a more fundamental point of view, the revealed mechanism offers a path to focus kinetic liquid energy through collective and resonant inertial cavitation bubble dynamics. 

\begin{acknowledgments}
Funding by the German Research Foundation (DFG Project No. 440395856) and the German Federal Ministry of Education and Research (Research Campus STIMULATE No. 13GW0473A). The authors gratefully acknowledge the Gauss Centre for Supercomputing e.V. for funding this project by providing computing time on the GCS Supercomputer SuperMUC-NG at Leibniz-Rechenzentrum. J.M.G. thanks project PID2020-115655GB-C21 from the Spanish MCIN/ AEI/10.13039/501100011033.
\end{acknowledgments}

\bibliography{references}

\end{document}


\preprint{AIP/123-QED}

\title[Supplementary: Jetting from resonant inertial cavitation]{Supplementary: Amplification of supersonic micro-jets by resonant inertial cavitation pair} 
\author{Yuzhe Fan}
\affiliation{Faculty of Natural Sciences, Institute for Physics, Department Soft Matter, Otto-von-Guericke University Magdeburg,39106 Magdeburg, Germany}
\affiliation{Research Campus STIMULATE, University of Magdeburg, Otto-Hahn-Straße 2, 39106 Magdeburg, Germany}
\author{Alexander Bußmann}
\affiliation{Chair of Aerodynamics and Fluid Mechanics, TUM School of Engineering and Design, Technical University of Munich, 85748 Garching bei München, Germany}
\affiliation{Munich Institute of Integrated Materials, Energy and Process Engineering (MEP), Technical University of Munich, 85748 Garching bei München, Germany}
\author{Fabian Reuter}
\affiliation{Faculty of Natural Sciences, Institute for Physics, Department Soft Matter, Otto-von-Guericke University Magdeburg,39106 Magdeburg, Germany}
\author{Hengzhu Bao}
 \affiliation{Suzhou University of Science and Technology, School of Physical Science and Technology, Suzhou 215009, China}
\author{Stefan Adami}
\affiliation{Munich Institute of Integrated Materials, Energy and Process Engineering (MEP), Technical University of Munich, 85748 Garching bei München, Germany}
\author{José M. Gordillo}
\affiliation{Área de Mecánica de Fluidos, Departamento de Ingenería Aeroespecial y Mecánica de Fluidos, Universidad de Sevilla, 41092 Sevilla, Spain}
\author{Nikolaus Adams}
\affiliation{Chair of Aerodynamics and Fluid Mechanics, TUM School of Engineering and Design, Technical University of Munich, 85748 Garching bei München, Germany}
\affiliation{Munich Institute of Integrated Materials, Energy and Process Engineering (MEP), Technical University of Munich, 85748 Garching bei München, Germany}
\author{Claus-Dieter Ohl}
\affiliation{Faculty of Natural Sciences, Institute for Physics, Department Soft Matter, Otto-von-Guericke University Magdeburg,39106 Magdeburg, Germany}
\affiliation{Research Campus STIMULATE, University of Magdeburg, Otto-Hahn-Straße 2, 39106 Magdeburg, Germany}
\date{\today}

\maketitle
\onecolumngrid

\section{Estimation of the maximum radius}
\begin{figure*}[h]
  \centering
  \includegraphics[]{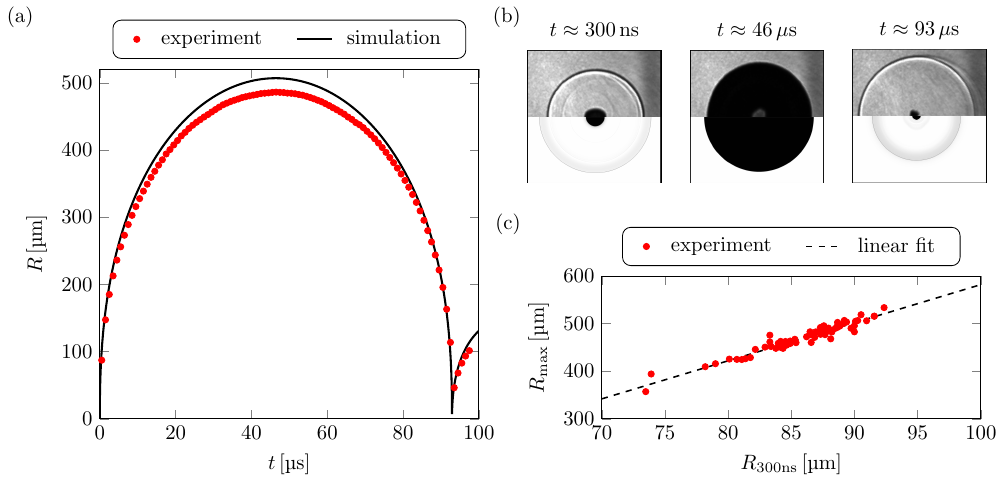}
  \caption{\textbf{Comparison between simulation and experiment for the dynamics of a single laser-induced cavitation bubble in water.} (a) Temporal evolution of the equivalent radius, $R(t)=(3/(4\pi)V(t))^{1/3}$ of the cavitation bubble, where $V(t)$ is the bubble volume. (b) Comparison of high-speed images of the experiment (upper tiles) and numerical Schlieren of the simulation (lower tiles) for three instances. At $t\approx \SI{300}{\nano\second}$, the bubble shape is shown shortly after seeding with its shock wave emission. $t\approx \SI{46}{\micro\second}$ shows the maximum expansion, and $t\approx \SI{93}{\micro\second}$ shows the bubble just after its collapse in the early rebounding stage with the collapse shock wave. (c) Relation between the maximum equivalent bubble radius, $R_\mathrm{max}$, and the bubble radius at $t\approx \SI{300}{\nano\second}$ after seeding, $R_{300\si{\nano\second}}$, for different experimental runs. The dashed line gives a linear fit. }
  \label{fig:single_bubble_comparison}
\end{figure*}

The maximum radius of the second bubble $R_{2,\mathrm{max}}$ is defined in a far-field situation, and during the bubble pair interaction, the second bubble is strongly deformed. Hence, it would reach a different volume in the single bubble case with the same amount of laser-deposited energy. However, it is reasonable to assume that, within the first $\SI{300}{\nano\second}$, the initial growth of the second bubble is not significantly altered with/without the existence of the first bubble due to their distance. Therefore, we study a number of 100 single bubble cases in water with a bubble radius of $R_\mathrm{max}=\SI{482}{\micro\meter}$ (median) in a far-field situation. Fig.~\ref{fig:single_bubble_comparison}a gives the radius as a function of time, and Fig.~\ref{fig:single_bubble_comparison}b shows three snapshots of the bubble dynamics. 
We thus establish a relation between the dynamics of the initial growth of the bubble and its maximum radius in the single bubble case. This relation is then used in the dual bubble configuration to predict the maximum radius of the second bubble it would take without the first bubble. The sought relation is given in Fig.~\ref{fig:single_bubble_comparison}c. We calculate the exact seeding time from the radius of the circular shock front to increase the temporal resolution beyond the imaging inter-frame period of the camera. We also use the shock wave fronts to read the respective bubble initiation positions, which the respective centers of the circular shock fronts give. The speed of the shock wave is assumed to be consistent with the sound speed.

\section{Numerical benchmark comparison with single bubble dynamics}
We use the dynamics of a single bubble to benchmark the numerical simulation. Figure~\ref{fig:single_bubble_comparison}a-b shows the radial oscillation of a single laser-induced bubble in our setup measured from high-speed imaging and obtained from the simulation. There is an excellent agreement between numerical and experimental data for the bubble lifetime, i.e., the time from initiation to collapse. The maximum radii differ by approximately $5\,\%$. We attribute the difference in bubble sizes to different modeling and geometrical effects. The bubble content is a mixture of water vapor and non-condensable gas, which we model using an ideal gas. Besides, the bubble geometry at the beginning slightly differs between experiment and simulation. In the experiment initially, the laser-induced plasma has an ellipsoidal shape, which we do not account for in the simulation, where we start from a spherical bubble nucleus (see $t\approx\SI{300}{\nano\second}$ in Fig.~\ref{fig:single_bubble_comparison}b). Further, in the experimental case, there are shock wave reflections at the cuvette walls and the water surface. Such effects are damped in the simulations by coarsening the mesh towards the boundaries and choosing a sufficiently large three-dimensional axisymmetric domain resembling unbounded cavitation bubble dynamics. 

\section{Neck collapse and jet ejection}
\begin{figure*}
  \centering
  \includegraphics[width=\textwidth]{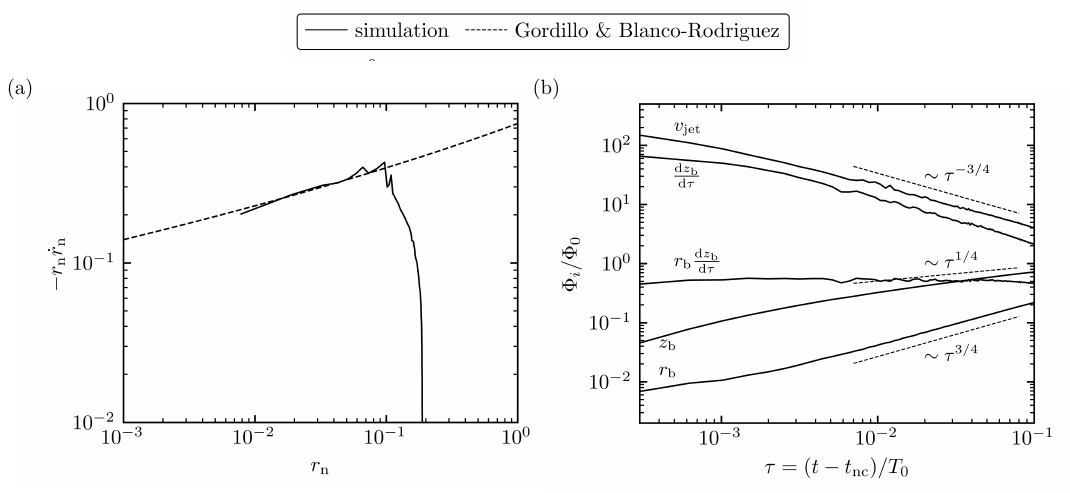}
  \caption{\textbf{Comparison of neck collapse and jet ejection between simulation and theory.} (a) Non-dimensional flow rate at the neck apex, $r_\mathrm{n}\dot{r}_\mathrm{n}$, over the neck radius, $r_\mathrm{n}$, during the collapse of the parabolic neck. (b) Non-dimensional jet characteristics of the radial jet base, $r_\mathrm{b}$, the axial jet base, $z_\mathrm{b}$, the jet flow rate, $r_\mathrm{b}\mathrm{d}z_\mathrm{b}/\mathrm{d}\tau$, and the jet base velocity, $v_\mathrm{jet}$. In both plots, the solid line corresponds to the simulation, and the dashed line represents the theoretical predictions of \citet{gordillo2023jets, gordillo2023jetsSupplementary}.}
  \label{fig:analytical_comparison}
\end{figure*}

When we compare the evolution of the parabolic neck with the analytical model of \citet{gordillo2023jets, gordillo2023jetsSupplementary}, we observe excellent agreement (Fig.~\ref{fig:analytical_comparison}). As a complementary study of the neck radius over time in the main manuscript, Fig.~\ref{fig:analytical_comparison}a, shows the non-dimensional radial flow rate per unit length of the neck, $r_\mathrm{n}\dot{r}_\mathrm{n}$, as a function of the neck radius, $r_\mathrm{n}$. During the neck growth, $r_\mathrm{n}>0.1$, the flow rate increases due to the surrounding pressure field induced by the collapsing torus bubble. Shortly after the torus collapse, $r_\mathrm{n}\approx0.1$, the flow rate reaches its maximum value, initiating the collapse of the parabolic neck. During this initial transient phase after the torus collapse, shock wave reflections/transmission on/through the neck surface lead to flow rate oscillations between $0.05<r_\mathrm{n}<0.13$. Once the inflow conditions reach a predominant radial profile, the analytical model of \citet{gordillo2023jetsSupplementary} (Eq. (2) and (7)) greatly predicts the collapse of the parabolic neck, where we use as initial conditions $r_\mathrm{n}(0)=0.073$, $r_\mathrm{n}\dot{r}_\mathrm{n}(0)=-0.365$, and $(r_\mathrm{n}/r_\mathrm{c})(0)=0.018$. Hence, our results show that the radial flow rate per unit length decreases logarithmically as the neck approaches the singularity on the axis of symmetry. This slow decrease finally fixes the flow rate per unit length, $q_\infty$, driving the ejection of the jet by mass conservation\cite{gordillo2023jets}.

After the neck radially converges on the axis of symmetry, the ultra-fast needle jet is ejected in the axial direction. In Fig.~\ref{fig:analytical_comparison}b, we show the evolution of the non-dimensional jet base characteristics over time from the simulations as a complementary study to the comparisons in the main manuscript. Shown are the radial jet width, $r_\mathrm{b}$, the axial jet position, $z_\mathrm{b}$, the axial jet base velocity, $\mathrm{d}z_\mathrm{b}/\mathrm{d}\tau$, and the jet velocity, $v_\mathrm{jet}$. All quantities are shown in non-dimensional form, where the reference length, velocity, and time are respectively $R_0=\SI{500}{\micro\meter}$, $V_0=(p_\infty/\rho_\infty)^{1/2}\approx\SI{10}{\meter\per\second}$, and $T_0=R_0/V_0=\SI{50}{\micro\second}$. 

We observe that, after a very short transient, taking place at a time scale $\sim\SI{50}{\nano\second}$, the flow rate per unit length, $q_\infty\propto r_\mathrm{b}\mathrm{d}z_\mathrm{b}/\mathrm{d}\tau$, driving the jet ejection, is fixed after the collapse of the neck and remains constant over the entire ejection process. Indeed, \citet{gordillo2023jets, gordillo2023jetsSupplementary} show that with this fixed driving flow rate, the temporal evolution of the jets produced after the implosion of parabolic cavities is given by $r_\mathrm{b}\propto \left(q^3_\infty(t-t_\mathrm{nc})^3/r^2_\mathrm{c}\right)^{1/4}$,
$z_\mathrm{b}\propto \left(q_\infty(t-t_\mathrm{nc}) r^2_\mathrm{c}\right)^{1/4}$, and 
$v_\mathrm{jet}\propto 2 \mathrm{d} z_\mathrm{b}/\mathrm{d} t\propto \left(q_\infty r^2_\mathrm{c}/(t-t_\mathrm{nc})^3\right)^{1/4}$, which is in excellent agreement with the numerical results in Fig.~\ref{fig:analytical_comparison}b. 

\section{Classification jetting regimes of anti-phase bubble pair}
\begin{figure*}
  \centering
  \includegraphics[width=0.98\textwidth]{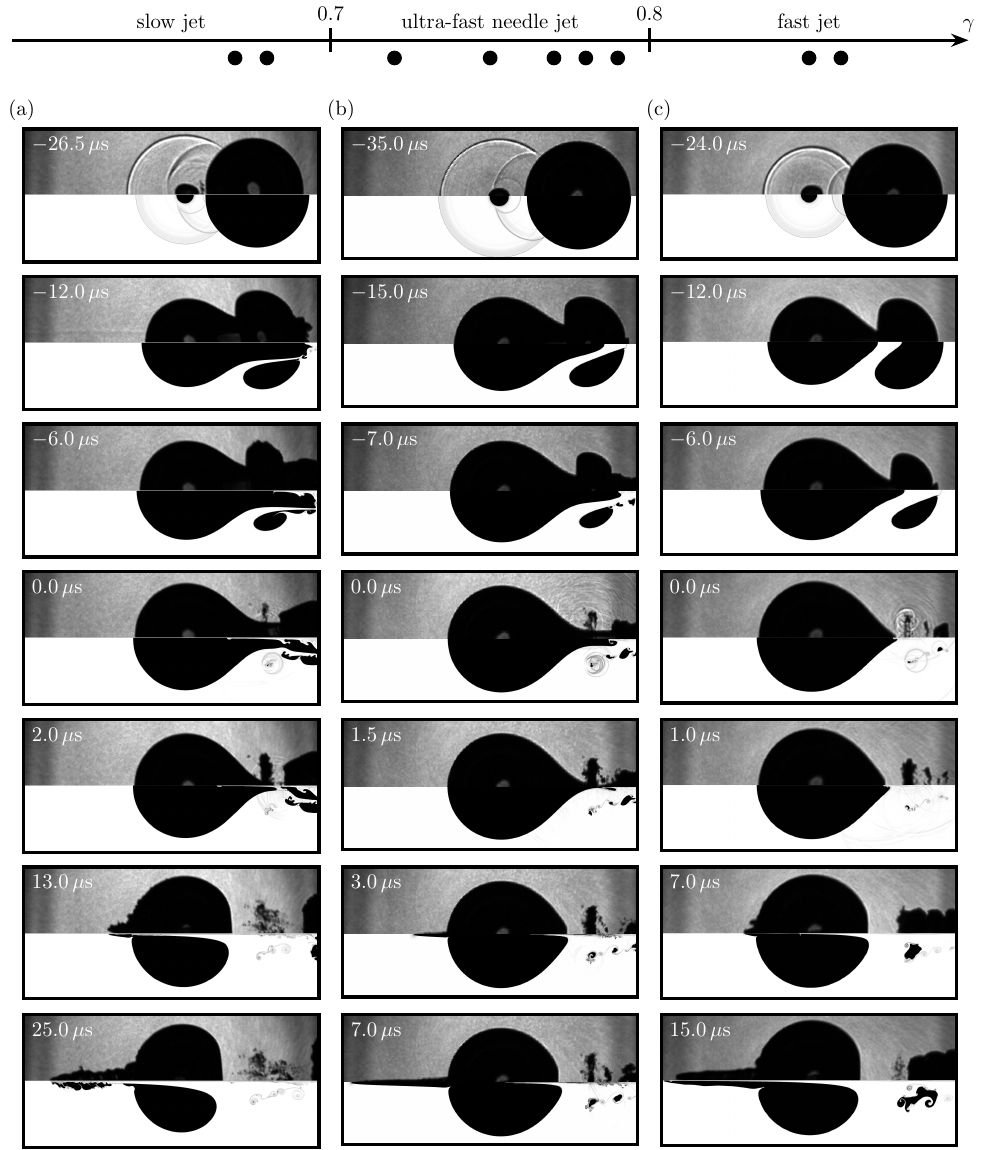}
  \caption{\textbf{Comparison of bubble jetting dynamics for different standoff distances.} Shown are the characteristic evolution of the three identified jetting regimes: slow jet ($\gamma=0.67$), ultra-fast needle jet ($0.7<\gamma<0.8$), and fast jet ($\gamma=0.83$). Each time instance compares the experiment (upper tile) and the simulation (lower tile). The experimental time instances are given in the upper left corner of each image. The considered experimental runs are indicated at the top of the image. (a) Slow jet regime for a standoff distance $\gamma\approx0.67$, (b) ultra-fast needle jet regime for a standoff distance $\gamma\approx0.75$, and (c) fast jet regime for a standoff distance $\gamma\approx0.83$.}
  \label{fig:standoff_distance_comparison}
\end{figure*}

\begin{figure*}
  \centering
  \includegraphics[width=0.98\textwidth]{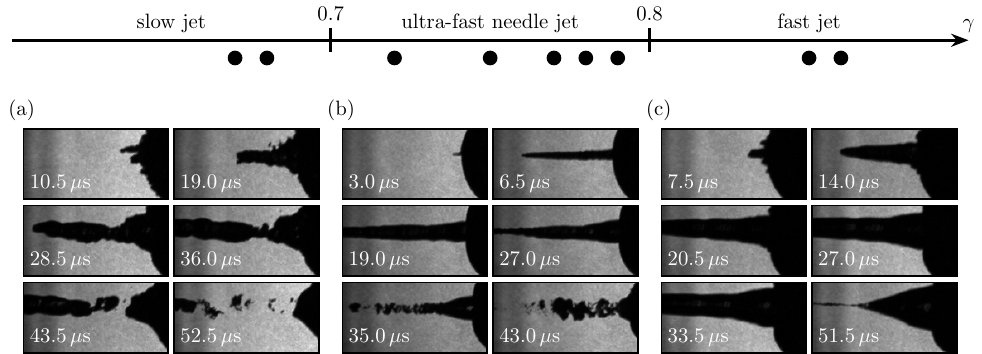}
  \caption{\textbf{Long time jet dynamics for different standoff distances.} Shown are the late-stage jet dynamics of the three identified jetting regimes: slow jet ($\gamma=0.67$), ultra-fast needle jet ($0.7<\gamma<0.8$), and fast jet ($\gamma=0.83$). Each time instance shows a zoom on the distal part of the second initiated bubble in the experiments. The time instances are given in the lower left corner of each image. The considered experimental runs are indicated at the top of the image. (a) Slow jet regime for a standoff distance $\gamma\approx0.67$, (b) ultra-fast needle jet regime for a standoff distance $\gamma\approx0.75$, and (c) fast jet regime for a standoff distance $\gamma\approx0.83$.}
  \label{fig:long_time_jet_dynamics}
\end{figure*}

In experiment and simulation, the ultra-fast needle jet in bubble pair anti-phase configuration only occurs consistently in a narrow range of standoff distances, $0.7<\gamma<0.8$. Figure~\ref{fig:standoff_distance_comparison} shows the dynamics for two standoff distances slightly outside the needle jet regime, i.e., $\gamma=0.67$ (left column) and $\gamma=0.83$ (right column). 
For the smaller standoff distance, the interaction between the two bubbles is stronger, and the neck formation is enhanced, leading to a larger and thicker neck ($t=-\SI{12}{\micro\second}$ to $t=-\SI{6}{\micro\second}$ in Fig.~\ref{fig:standoff_distance_comparison}a). The rather corrugated neck topology leads to more complex transmission/reflection patterns of the torus shock wave through/from the neck surface. Hence, the neck breakup does not follow the parabolic collapse. This complex collapse leads to more significant partitioning of the collapse energy, reducing the overall axial jet speed by a factor of four to five compared to the ultra-fast needle jet ($t=\SI{13}{\micro\second}$ to $t=\SI{25}{\micro\second}$ in Fig.~\ref{fig:standoff_distance_comparison}a). 
For the larger standoff distance, the interaction between the bubbles is reduced such that the neck does not grow as long as in the ultra-fast needle jet ($t=-\SI{12}{\micro\second}$ to $t=-\SI{6}{\micro\second}$ in Fig.~\ref{fig:standoff_distance_comparison}c). Hence, the torus shock wave does not impact the neck surface but directly on the axis of symmetry. Hence, the liquid flow focusing geometry of the neck is not realized (($t=-\SI{0}{\micro\second}$ in Fig.~\ref{fig:standoff_distance_comparison}c). Compared to the slow jet regime at $\gamma=0.67$ an energy focusing on the axis of symmetry still occurs by the radial convergence of the toroidal shock wave, resulting in a stronger pressure gradient on the symmetry axis. This singularity leads to the ejection of faster jet speeds compared to $\gamma=0.67$. However, the jet speeds are slower by a factor of two to three than in the ultra-fast needle jet regime, where liquid flow focusing by the collapse of the parabolic neck is the relevant phenomenon. 
The different energy partitioning effects also translate into the topology of the long-time jet dynamics (Fig.~\ref{fig:long_time_jet_dynamics}). For the smallest standoff distance, $\gamma=0.67$, the jet topology shows significant asymmetries and undulations after the jet pierces the distal part of the second bubble ((Fig.~\ref{fig:long_time_jet_dynamics}a). When strong pressure gradients on the axis of symmetry cause the jet ejection, the jet topology is smoother ((Fig.~\ref{fig:long_time_jet_dynamics}b-c).

\newpage
\bibliography{references}